\documentclass[aps,pra,twocolumn]{revtex4}
\usepackage[pctex32]{graphics}

\begin{document}
\title{Engineering Schr\"{o}dinger cat states through cavity-assisted interaction of
coherent optical pulses}
\author{B. Wang and L.-M. Duan}
\address{FOCUS center and MCTP, Department of Physics, University of Michigan, Ann
Arbor, MI 48109}

\begin{abstract}
We propose a scheme to engineer Schr\"{o}dinger-cat states of
propagating optical pulses. Multi-dimensional and multi-partite
cat states can be generated simply by reflecting coherent optical
pulses successively from a single-atom cavity. The influences of
various sources of noise, including the atomic spontaneous
emission and the pulse shape distortion, are characterized through
detailed numerical simulation, which demonstrates practicality of
this scheme within the reach of current experimental technology.

\end{abstract}
\maketitle Schr\"{o}dinger-cat states, as a distinct class of
non-classical states, have attracted extensive research interests
recently in the contexts of fundamental test of quantum mechanics
and quantum information theory. For bosonic modes, the cat states
are typically referred as quantum superpositions of coherent
states. Such states are of critical importance for investigation
of the decoherence process and the quantum-classical boundary
\cite {1}, for fundamental test of the quantum nonlocality \cite
{2}, and for implementation of quantum computation and
communication \cite {3}. Significant experimental efforts have
been made for realization of such cat states in different physical
systems \cite {1,4}. Till now, such states have been successfully
generated for phonon modes of a single trapped ion \cite {1}, or
for a microwave mode confined inside a superconducting cavity
\cite {4}.

There are also great interests in generating Schr\"{o}dinger-cat
states for propagating optical pulses. The motivation is at least
two-folds: firstly, for applications in test of quantum
nonlocality or for quantum computation and communication, one
needs to use cat states of propagating optical pulses; Secondly,
for propagating pulses, with assistance of linear optical devices
(such as a beam splitter), it is possible to generate a larger
class of cat states targeted to different kinds of applications
\cite {5}. The proposals for generating cat states of optical
pulses are typically based on either the Kerr nonlinearity or
postselections from the non-linear detectors \cite {6,7,8}.
Although the Kerr nonlinearity in principle provides a method for
deterministic generation of the cat states, it is well known that
such nonlinearity in typical materials is too small to allow cat
state generation from weak coherent pulses.

In this paper, we propose a scheme to engineer Schr\"{o}dinger-cat
states for propagating optical pulses based on the
state-of-the-art of the cavity technology. A single atom has been
trapped inside a strong-coupling cavity with seconds of trapping
time \cite {9,10,11}. With such a setup, we propose to generate a
large class of cat states simply by successively reflecting weak
coherent pulses from a cavity mirror. Together with a few beam
splitters, we can generate multi-partite and multi-dimensional cat
states, and preparation of such states is a necessary step for
several distinct applications, such as for loop-hole-free
detection of Bell inequalities with homodyne detections \cite
{12}, or for quantum coding and computation \cite {13}. This
scheme also extends an earlier photonic quantum computation
proposal by Duan and Kimble \cite {14} to the continuous variable
regime, eliminating the requirement of single-photon pulses as the
cavity inputs. To characterize the influences of various sources
of experimental noise on this scheme, we develop a numerical
simulation method which can be used to calculate how large cat
states one can produce, the effects of the atomic spontaneous
emission and the pulse shape distortion, etc. The calculation
shows that substantial cat states can be generated within the
reach of the current technology.

\begin{figure} [tbp]
\includegraphics{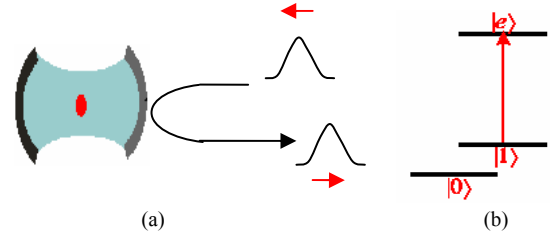}\caption[Fig.1]{(a)
Schematic setup for generation of cat states by reflecting a
coherent optical pulse from a single-atom cavity. (b) The relevant
level structure of the atom trapped in the cavity.}
\end{figure}

First, let us briefly introduce the basic idea of this scheme. We
consider an atom of three relevant levels trapped inside an
optical cavity. The level configuration is shown in Fig.1, where
$|0\rangle $ and $|1\rangle $ are levels in the ground-state
manifold with different hyperfine spins. The transition from the
level $|1\rangle $ to $|e\rangle $ is resonantly coupled to a
cavity mode $a_{c}$, which is resonantly driven by an input
optical pulse prepared in a weak coherent state $\left| \alpha
\right\rangle $. The level $|0\rangle $\ is de-coupled from the
cavity mode due to the large
detuning from the hyperfine frequency. If the atom is prepared in the level $%
|0\rangle $, the input pulse is resonant with the bare cavity mode
$a_{c},$ and after resonant reflection it will acquire a phase of
$e^{i\pi }$ from standard quantum optics calculation \cite {15}.
The effective state of the pulse is then given by $\left| -\alpha
\right\rangle $. However, if the atom is prepared in the level
$|1\rangle $, due to the strong atom-cavity coupling, the
frequency of the dressed cavity mode is significantly detuned from
that of the input pulse. In this case, one would expect
intuitively that the cavity mode structure does not play an
important role here, and the reflection is then\ similar to the
reflection from a mirror, which keeps the
pulse shape and phase unchanged. The effective pulse state remains to be $%
\left| \alpha \right\rangle $. Our later calculation will confirm this
expectation if the input amplitude $\alpha $ is within a reasonable range.

In order to generate a Schr\"{o}dinger-cat state, we simply
prepare the
trapped atom in a superposition state $\left( |0\rangle +|1\rangle \right) /%
\sqrt{2}$. Then, if we reflect a coherent pulse $\left| \alpha \right\rangle
$, following the above analysis, the final atom-photon state becomes an
entangled cat state
\begin{equation}
|\Psi _{c}\rangle =\left( |0\rangle \left| -\alpha \right\rangle +|1\rangle
\left| \alpha \right\rangle \right) /\sqrt{2}.
\end{equation}
This entangled cat state can be experimentally verified though a
homodyne detection of the state of the reflected photon pulse
\cite {16}, correlated with a measurement of the atomic state in
the basis $\left\{ |\pm \rangle =\left( |0\rangle \pm |1\rangle
\right) /\sqrt{2}\right\} $. With homodyne detections, one can
also measure the Wigner function of the optical field though
quantum state tomography \cite {17}, which fully characterizes the
non-classicality of the cat state.

With some extensions to the above method, we can generate more
complicate types of cat states. Firstly, by bouncing a series of
coherent pulses (say, $n$ pulses), each initially in the state
$\left| \alpha \right\rangle $, successively from the same
single-atom cavity, one will get the state $\left( |0\rangle
\left| -\alpha \right\rangle ^{\otimes n}+|1\rangle \left| \alpha
\right\rangle ^{\otimes n}\right) /\sqrt{2}$, which yields
entangled multi-partite cat states $\left( \left| -\alpha
\right\rangle ^{\otimes n}\pm \left| \alpha \right\rangle
^{\otimes n}\right) $ (unnormalized) for the pulses after a
measurement of the atomic state in the basis $\left\{ |\pm \rangle
\right\} $. Secondly, after generation of the state $\left( \left|
-\alpha \right\rangle +\left| \alpha \right\rangle \right) $ for
the pulse, one can transfer it to the state $\left( \left| \alpha
\right\rangle +\left| 3\alpha \right\rangle \right) $ through a
simple lineal optical manipulation (for instance, by interfering
this pulse with another phase-locked stronger laser pulse at an
unbalanced beam splitter, one can shift up each coherent component
of the cat state by an amplitude of $2\alpha $). Then, if we
reflect this pulse again from the same cavity, we will get a state
$\left( \left| -3\alpha \right\rangle +\left| -\alpha
\right\rangle +\left| \alpha \right\rangle +\left| 3\alpha
\right\rangle \right) $ for the pulse conditioned on that a
measurement of the atom gives the $|+\rangle $ state. It is
straightforward to extend this idea to generate the
multi-dimensional
cat states $\sum_{i=-n}^{n+1}\left| \left( 2i-1\right) \alpha \right\rangle $%
, and such kind of states have important applications for
continuous-variable quantum coding \cite {13} and loop-hole-free
detection of the Bell inequalities with efficient homodyne
measurements \cite {12}.

In the above, we have given the basic idea for preparation of the
cat state (1) and described its various extensions. To have real
understanding and characterization of this process, however, we
need more detailed theoretical modeling of the interaction between
the cavity atom and the light pulse. First, we want to know how
large a cat state one can prepare. If the amplitude $|\alpha|$ is
large, one would expect that a single atom can not significantly
change the property of a strong pulse, therefore the output state
would be different from the state described by Eq.(1). Second, in
reality there are always experimental noises, such as the photon
loss due to the atomic spontaneous emission and the mirror
scattering, the inherent pulse shape distortion after reflection
from the cavity, and the random variation of the atom-photon
coupling rate caused by the thermal atomic motion. One needs to
characterize the influence of these sources of noise on the
generation of the cat state.

The input to the cavity is a coherent optical pulse, whose state $|\alpha
\rangle _{in}$ can be described by $|\alpha \rangle _{in}=\exp \left[ \alpha
\int_{0}^{T}f_{in}(t)a_{in}^{\dagger }(t)dt\right] |$vac$\rangle $, where $%
a_{in}(t)$ is a one-dimensional quantum field operator with the standard
commutation relation $[a_{in}(t),a_{in}(t^{\prime })]=\delta (t-t^{\prime })$,
$f_{in}(t)$ describes the input pulse shape with the normalization $%
\int_{0}^{T}\left| f_{in}(t)\right| ^{2}dt=1$\ ($T$ is the pulse duration),
and $|$vac$\rangle $ represents the vacuum state for all the optical modes.
The average photon number of the pulse is given by $\left| \alpha \right| ^{2}$%
. The input pulse drives the cavity mode $a_{c}$ through the
Langevin equation \cite {15}

\begin{equation}
\dot{a}_{c}=-i[a_{c},H]-\frac{\kappa }{2}a_{c}-\sqrt{\kappa }a_{in}(t),
\end{equation}
where $\kappa $ is the cavity decay rate, and the Hamiltonian $H$ describes
the atom-cavity interaction with the form
\begin{equation}
H=\hbar g\left( |e\rangle \langle 1|a_c+|1\rangle \langle
e|a^{\dag }_c\right) .
\end{equation}
Here, $g$ is the atom-cavity coupling rate. The cavity output
field $a_{out}$ is connected to the input through the input-output
relation
\begin{equation}
a_{out}(t)=a_{in}(t)+\sqrt{\kappa }a_{c}(t).
\end{equation}

We need to find out the quantum state of the cavity output field
$a_{out}$ through the series of equations (2)-(4). As they are
nonlinear operator equations with infinite modes, it is hard to
solve them even numerically. For the case of a single-photon pulse
input, a numerical method has been developed in Refs. [14,18]
based on the mode discretization and expansion. But that method
does not work if the photon number of the input could be larger
than $1$, as it is the case for the present work. To attack this
latter problem, we propose a variational
method based on the following observation: if the atom is in the state $%
|0\rangle $, the Hamiltonian (3) does not play a role, and Eqs.
(2) and (4) become linear, from which we observe that the state
$|\phi _{0}\rangle _{out} $ of the output field can be exactly
written as $|\phi _{0}\rangle _{out}=\exp \left[ \alpha
\int_{0}^{T}f_{out}^{\left( 0\right) }(t)a_{out}^{\dagger
}(t)dt\right] |$vac$\rangle $. The normalized shape function can
be expressed as $f_{out}^{\left( 0\right)
}(t)=-\int\frac{\frac{\kappa}{2}+i\omega}{\frac{\kappa}{2}-i\omega}
\exp\left[i{\omega}t\right]f_{in}(\omega)d\omega $, where
$f_{in}(\omega)$ is the Fourier transform of $f_{in}(t)$. The
output optical field is still in an effective single-mode coherent
state, but with the mode shape function $f_{out}^{\left( 0\right)
}(t)$ generally different from the input shape $f_{in}(t)$. If the
atom is in the state $|1\rangle $, it is reasonable to make the
ansatz that the output optical field is also in an effective
single-mode coherent state $|\phi _{1}\rangle _{out}=\exp \left[
\alpha _{1}\int_{0}^{T}f_{out}^{\left( 1\right) }(t)a_{out}^{\dagger }(t)dt%
\right] |$vac$\rangle $, but with probably a different normalized mode shape
function $f_{out}^{\left( 1\right) }(t)$. In general, the amplitude $%
\alpha _{1}$ can be different from $\alpha $ (actually $\left|
\alpha _{1}\right| ^{2}<\left| \alpha \right| ^{2}$) because of
the atomic spontaneous emission loss. Due to that loss, some of
the photons are scattered to other directions, so we have a weaker
output field. To find out the functional form of
$f_{out}^{\left(1\right)}(t)$, we note that under the above
ansatz, the expectation value of the input-output equation (4)
leads to
\begin{equation}
\alpha _{1}f_{out}^{\left( 1\right) }(t)=\alpha f_{in}(t)+\sqrt{\kappa }%
\left\langle a_{c}(t)\right\rangle .
\end{equation}
The expectation value of the cavity mode operator $a_{c}(t)$ can be found by
solving the corresponding master equation for the atom-cavity-mode density
operator $\rho $%
\begin{eqnarray}
\dot{\rho} &=&-\frac{i}{\hbar }[H_{eff},\rho ]+\frac{\kappa
}{2}\left( 2a_{c}\rho a_{c}^{\dagger }-a_{c}^{\dagger }a_{c}\rho
-\rho a_{c}^{\dagger
}a_{c}\right)   \nonumber \\
&+&\frac{\gamma _{s}}{2}\left( 2\sigma _{-}\rho \sigma _{+}-\sigma
_{+}\sigma _{-}\rho -\rho \sigma _{+}\sigma _{-}\right) ,
\end{eqnarray}
where $\sigma _{-}=|1\rangle \langle e|$ and $\sigma _{+}=|e\rangle \langle
1|$ are the atomic lowering and raising operators, and the effective
Hamiltonian $H_{eff}=\hbar (g\sigma _{+}a+i\sqrt{\kappa }\langle
a_{in}\rangle a)+H.c.$ Comparing with the Hamiltonian (3), we add the term $%
i\hbar \sqrt{\kappa }\langle a_{in}\rangle a+H.c.$ to account for
the driving from the input pulse. After that correction, the
cavity decay and the atomic spontaneous emission loss can then be
described by the last two terms of the master equation (6), where
$\gamma _{s}$ denotes the spontaneous emission rate. The density
operator $\rho $ can be solved from the master equation (6) with
the standard numerical method, from which we get the expectation
value $\left\langle a_{c}(t)\right\rangle =tr\left( \rho
a_{c}(t)\right) $. Then, following Eq. (5), we can determine the
output amplitude $\alpha _{1}$ and its pulse shape
$f_{out}^{\left( 1\right) }(t)$.
\begin{figure}[tbp]
\includegraphics{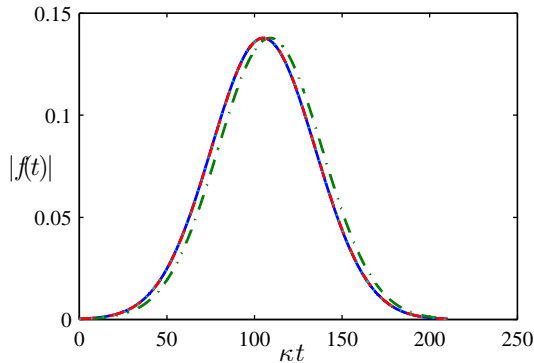}\caption[Fig.2]{Pulse shape functions for the input and output pulses.
The solid curve shows the shape of input pulse. The
dash-dotted, dashed, and dotted curves correspond to the output pulses with $%
g=0$(for the atom in the level $|0\rangle )$, $g/\protect\kappa =3$, and $g/\protect%
\kappa =6$, respectively. In the calculation, we assumed $\protect\gamma %
=\protect\kappa $.}
\end{figure}

With the above method, we calculate the pulse shapes
$f_{out}^{\left( 0\right) }(t)$ and $f_{out}^{\left( 1\right)
}(t)$ of the output optical field with the atom in the state
$|0\rangle $ and $|1\rangle $, respectively. For this calculation,
we take a Gaussian shape for the input pulse with
$f_{in}(t)\propto \exp \left[ -(t-T/2)^{2}/(T/5)^{2}\right] ,$
where $T$ characterizes the pulse duration. The results are shown
in Fig. 2, which demonstrates that the shape functions
$f_{out}^{\left( 0\right) }(t)$ and $f_{out}^{\left( 1\right)
}(t)$ of the output pulses
overlap very well with $f_{in}(t)$ of the input pulse if the pulse duration $%
T\gg 1/\kappa $. Furthermore, the global phase factors of $f_{out}^{\left(
0\right) }(t)$ and $f_{out}^{\left( 1\right) }(t)$ are given by $-1$ and $1$%
, respectively, which confirms our previous expectation: if the
atom is initially prepared in a superposition state $\left(
|0\rangle +|1\rangle \right) /\sqrt{2}$, the final atom-photon
state will be the desired entangled cat state $|\Psi _{c}\rangle $
as shown in Eq. (1), where $\left| -\alpha \right\rangle $ and
$\left| \alpha \right\rangle $ are coherent states of the output
mode with the mode shape function $-f_{out}^{\left( 0\right)
}(t)\approx f_{out}^{\left( 1\right) }(t)\approx f_{in}(t)$. In
the same figure, we have also shown the output shape
$f_{out}^{\left( 1\right) }(t)$ for different atom-photon coupling
rates $g$. Both the phase and the amplitude of $f_{out}^{\left(
1\right) }(t)$ are very insensitive to random variation of $g$
within a certain range. For instance, even if $g$ varies by a
factor of $2$ from $6\kappa $ to $3\kappa $ (which is the typical
variation range of $g$ caused by the atomic thermal motion), the change in $%
f_{out}^{\left( 1\right) }(t)$ is negligible ($<10^{-4}$).

\begin{figure}[tbp]
\includegraphics{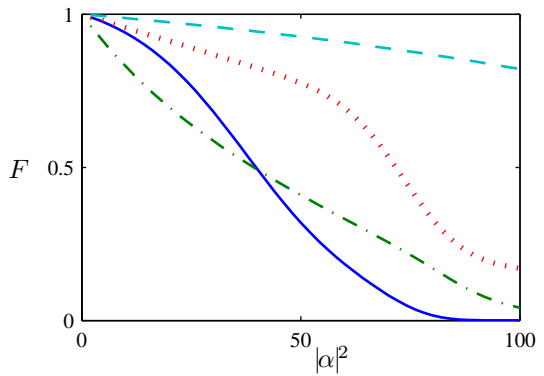}\caption[Fig.3]{The cat state Fidelity shown
as a function of the average input photon number $\left
|\alpha\right |^2$ when the spontaneous emission rate is set to
zero ($\protect\gamma_s =0$). Other parameters: $g/\protect\kappa
=3$ and $\protect\kappa T=210$ for the solid curve;
$g/\protect\kappa =6$ and $\protect\kappa T=210$ for the dotted
curve; $g/\protect\kappa =6$ and $\protect\kappa T=100$ for the
dash-dotted curve; $g/\protect\kappa =6$ and $\protect\kappa
T=400$ for the dashed curve. }
\end{figure}

To quantify the limit of the cat states that one can prepare and
the influence of some practical noise, we introduce several
quantities to measure the quality of the cat state preparation.
First, the distortion
between the output and the input pulses can be measured by their mismatching $%
\xi _{1}=1- \int f_{in}^{\ast }f_{out}^{\left( 1\right) }(t)dt$ and $%
\xi _{0}=1- \int f_{in}^{\ast }f_{out}^{\left( 0\right) }(t)dt$ (as $f_{out}^{\left( 0\right) }(t)$
has an opposite phase). With typical experimental parameters, $\xi _{0}\gg \xi _{1}$, so $%
\xi _{0}$ has the dominant contribution to the imperfection of our
scheme. Second, the effect
of the spontaneous emission loss can be quantified by the photon loss parameter $%
\eta =1-\left| \alpha _{1}\right| ^{2}/$ $\left| \alpha \right|
^{2}$, which represents the fraction of the photons scattered to
other directions instead of to the cavity output \cite {19}. Both
the pulse shape distortion and the photon loss have contributions
to the imperfection of the final cat state, which can be
characterized by the state fidelity. The ideal cat state is given
by $|\Psi _{c}\rangle $ in Eq. (1), while with noise, the
real state obtained is denoted by a density matrix $\rho _{%
%TCIMACRO{\func{real}}%
%BeginExpansion
\mathop{\rm real}%
%EndExpansion
}$. The fidelity, defined as $F\equiv \left\langle \Psi _{c}\right| \rho _{%
%TCIMACRO{\func{real}}%
%BeginExpansion
\mathop{\rm real}%
%EndExpansion
}|\Psi _{c}\rangle $, can be expressed by $\xi _{0}$ and $\eta $ as

\begin{equation}
F\approx \left|\frac{e^{-|\alpha |^{2}\left( 1-\sqrt{1-\eta
}\right) }+e^{-|\alpha |^{2}\xi _{0}}}{2}\right|^{2},
\end{equation}
where we neglect the contribution of $\xi _{1}$ as $\xi _{1}\ll
\xi _{0}$.
\begin{figure}[tbp]
\includegraphics{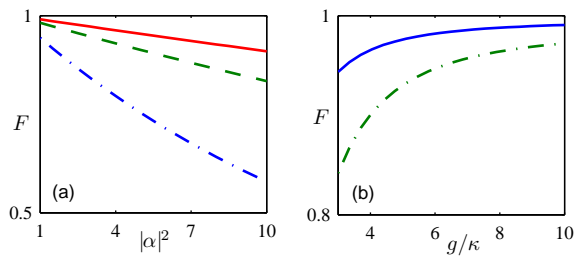}\caption[Fig.4]{(a)
The cat state fidelity shown as a function of the average photon
number $\alpha^2 $ of the input pulse. The dash-dotted, dashed
and solid curves correspond to $g/\protect\kappa =3$, $g/\protect\kappa =6$, and $g/\protect%
\kappa =10$, respectively. (b) The fidelity shown as a function of
the coupling rate $g$. The solid and dashed curves correspond to
the average input photon number $\left| \alpha _{1}\right| ^{2}=1$
and $\left| \alpha _{1}\right| ^{2}=3$,
 respectively. In both calculations for Figs. 4a and 4b, we have taken $%
\protect\gamma =\protect\kappa $ and $\protect\kappa T=210$. }
\end{figure}
First, let us examine the intrinsic limit to the amplitude of the
cat state that one can prepare even if we neglect the influence of
practical photon loss. This intrinsic limit comes from the fact
that a single cavity atom can not change much the property of a
strong optical field (i.e., with a large $\alpha$).
For that purpose, we simply set the spontaneous emission rate $\gamma _{s}=0$%
, and look at the state fidelity as a function of the cat amplitude $\alpha$%
. The result is shown in Fig. 3, with the maximum achievable cat amplitude $%
\alpha $ depending on the atom-cavity coupling rate $g$ and the
pulse duration $T$. In general, a larger $g$ and a longer $T$ help
for generation of a large cat state. In particular, the fidelity
increases dramatically when we increase the pulse duration $T$ as
the shape distortion parameter $\xi _{0}$ significantly reduces
for a narrower bandwidth pulse.

We then take the influence of practical noise into account, and
investigate under typical experimental configurations, how large a
cat state one can prepare. With the spontaneous emission rate
$\gamma _{s}=\kappa $, the state fidelity $F$ is shown as a
function of the cat amplitude in Fig. 4a, and as a function of the
coupling rate in Fig. 4b. The fidelity increases with the coupling
rate $g$ and decreases with the cat amplitude $\alpha $, as one
would expect. Under a reasonable atom-cavity coupling rate
$g\approx 10\kappa $ comparable with the current technology, a cat
state with a remarkable amplitude $\alpha \approx 3.4$
(corresponding to an entangled state of about $10$ photons) could
be generated with a $90\%$ fidelity.

In summary, we have proposed a scheme to generate and control
multi-partite and high-dimensional Schr\"{o}dinger-cat states for
propagating optical pulses. The scheme is based on the
state-of-the-art of the cavity technology. We have developed a
variational calculation method which can be used to solve
efficiently the interaction between the input-output quantum field
and the cavity atom. This calculation technique enables us to
quantitatively characterize the influence of various sources of
practical noise on the performance of this scheme.

We thank Jeff Kimble for helpful discussions. This work was
supported by the ARDA under ARO contract, the FOCUS seed funding,
and the A. P. Sloan Fellowship.


\begin{thebibliography}{99}
\bibitem {1} C. Monroe, D.M. Meekhof, B.E. King, D.J. Wineland, \textit{Science} \textbf{272}, 1131 (1996).
\bibitem {2} V. Buzek, A. Vidiella–Barranco, and P. L. Knight, \textit{Phys. Rev. A} \textbf{45}, 6570 (1992).
\bibitem {3} T. C. Ralph, A. Gilchrist, and G. J. Milburn, \textit{Phys. Rev. A} \textbf{68}, 042319 (2003).
\bibitem {4} M. Brune, E. Hagley, J. Dreyer, X. Maître, A. Maali, C. Wunderlich, J. M. Raimond, and S.
Haroche, \textit{Phys. Rev. Lett.} \textbf{77}, 4887 (1996).
\bibitem {5} S. J. van Enk, \textit{Phys. Rev. Lett.} \textbf{91}, 017902 (2003).
\bibitem {6} K. M. Gheri and H. Ritsch, \textit{Phys. Rev. A} \textbf{56}, 3187 (1997).
\bibitem {7} S. Song, C. M. Caves, and B. Yurke, \textit{Phys. Rev. A} \textbf{41}, 5261 (1990).
\bibitem {8} A. P. Lund, H. Jeong, T. C. Ralph, and M. S. Kim, \textit{Phys. Rev. A} \textbf{70}, 020101(R) (2004).
\bibitem {9} J. McKeever, A. Boca, A. D. Boozer, J. R. Buck, H. J. Kimble, \textit{Nature} \textbf{425}, 268 (2003).
\bibitem {10} J. McKeever, J. R. Buck, A. D. Boozer, A. Kuzmich, H.-C. Nägerl, D. M. Stamper-Kurn, and H. J.
Kimble, \textit{Phys. Rev. Lett.} \textbf{90}, 133602 (2003).
\bibitem {11} G. R. Guth\" {o}hrlein, M. Keller, K. Hayasaka, W. Lange, and H.
Walther, \textit{Nature} \textbf{414}, 49 (2001).
\bibitem {12} J. Wenger, M. Hafezi, F. Grosshans, R. Tualle-Brouri, and P.
Grangier, \textit{Phys. Rev. A} \textbf{67}, 012105 (2003).
\bibitem {13} D. Gottesman, A. Kitaev, J. Preskill, \textit{Phys.Rev. A} \textbf{64},
012310 (2001).
\bibitem {14} L.-M. Duan and H. J. Kimble, \textit{Phys. Rev. Lett.} \textbf{92}, 127902
(2004).
\bibitem {15} D. F. Walls, and G. J. Milburn, \textit{Quantum Optics} (Springer-Verlag, Berlin, 1994).
\bibitem {16} V. Buzek, and P. L. Knight, \textit{Progress in Optics}, vol. \textbf{XXXIV}, edited by E. Wolf (North Holland, Amsterdam 1995), and refs.
therein.
\bibitem {17} G. M. D'Ariano, M. F. Sacchi, and P. Kumar
\textit{Phys. Rev. A} \textbf{59}, 826 (1999).
\bibitem {18} L.-M. Duan, A. Kuzmich, H. J. Kimble, \textit{Phys. Rev. A}, \textbf{67}, 032305 (2003).
\bibitem{19}  Other sources of photon loss, such as the photon absorption
and scattering by the cavity mirrors, can be similarly described
by this quantity.
\end{thebibliography}
\end{document}